\documentclass[intlimits,twoside,a4paper]{article}
\usepackage{amsmath,amssymb}
\usepackage{graphicx}
\usepackage{wrapfig}
\usepackage[T2A]{fontenc}
\usepackage[cp1251]{inputenc}

\usepackage{threeparttable}

\usepackage[eqsecnum]{cmpj}

\issue{2011}{14}{4}{43601}

\doinumber{10.5488/CMP.14.43601}

\title[Effects of polymer concentration and  chain length  on aggregation]%
{Effects of polymer concentration  and  chain length \\ on aggregation in physically associating \\polymer solutions}

\author[X.-G. Han \textsl{et al.}]{X.-G.~Han\refaddr{label1}\thanks{E-mail: xghan0@163.com}\,,
 X.-F.~Zhang\refaddr{label1}, Y.-H.~Ma\refaddr{label1}, C.-X.~Zhang\refaddr{label2}, Y.-B.~Guan\refaddr{label2}}

\authorcopyright{X.-G.~Han, X.-F.~Zhang, Y.-H.~Ma, C.-X.~Zhang, Y.-B.~Guan, 2011}

\addresses{
\addr{label1} Inner Mongolia Key Laboratory for Utilization of Bayan Obo Multi-Metallic Resources: \\ Elected State Key Laboratory,
School of Mathematics, Physics and Biological
engineering, \\ Inner Mongolia University of Science and Technology,
014010~Baotou, China \addr{label2} Department of Physics, Jilin University,
130021~Changchun, China }

\date{Received July 2, 2011, in final form September 19, 2011}

\begin{document}

\maketitle

\begin{abstract}
The effects of  polymer concentration  and  chain length on
aggregation in associative polymer solutions, are studied using
self-consistent field lattice model. Only two inhomogenous
morphologies, i.e. microfluctuation homogenous (MFH) and micelle
morphologies, are observed in the systems with different chain
lengths. The temperatures at which the above two inhomogenous
morphologies first appear, which are denoted by $T_{\mathrm{MFH}}$ and $T_{\mathrm{m}}$,
respectively, are dependent on polymer concentration and chain
length. The variation of the logarithm of critical MFH concentration
with the logarithm of chain length fulfils a linear-fitting
relationship with a slope equaling $-1$. Furthermore, the variation of
the average volume fraction of stickers at the micellar core (AVFSM)
with polymer concentration and chain length is focused in the system
at $T_{\mathrm{m}}$. It is founded by calculations that the above behavior of
AVFSM, is explained in terms of intrachain and interchain
associations.

\pacs 61.25.Hp, 87.15.Nr, 82.60.Fa
\end{abstract}

\section{Introduction}

Physically, associating polymers are polymer chains containing a
small fraction of attractive groups along the backbones. The
attractive groups, for example, solvophobic groups, tend to form
physical links which can play an important role in reversible
junctions between different polymer chains. The junctions can be
broken and recombined frequently on experimental time scales. This
property of junctions makes associative polymer solutions behave reversibly
when ambient conditions, such as temperature and
concentrations, change. This tunable characteristic of the system
produces extensive applications~\cite{Clar1987,Slat1998,Tong2001,Tayl1998} that possess great potential
as smart materials~\cite{Brun2001,Gree2008,Cord2008}.

The attractive groups (also called sticker monomers) drive the
self-assembly of the polymers, leading to the formation of polymeric
micelles in physically associating polymer solutions (PAPSs). In
telechelic~\cite{Alam1996,Chas1997} and multiblock~\cite{Han2010}
associative polymers, flower micelles were observed. This
aggregation, as well as their ability to form bridges between
micelles, profoundly affect their macroscopic properties,
particularly their rheological behavior, which is required in
many applications. In telechelic associative polymers, the effects of
architectural parameters of polymers have been assessed, such as
chain length, end-group length, and chemical
composition~\cite{Chas1998,Beau2002,Beau2002a,Lafl2003,Lafl2003a}. In
multiblock associative polymers, the effect of chain
architecture of polymer on the property of
aggregates~\cite{Bala1991,Brow1992} was studied, which suggested that
chain architecture can be an important factor in controlling
macroscopic properties of the systems. However, the above studies of
multiblock PAPSs were carried out in two dimensions and at low
concentrations.

 It is well known that self-consistent field theory (SCFT), as a
mean-field theory, has been applied to the study of a great deal of
problems in polymeric
systems~\cite{Orland1996,Mats1994,Tang2004,He2004}. Recently, SCFT has been
applied to the study of the properties of micelles in polymer
solutions~\cite{Cava2006,Jeli2007,Char2008}. In the previous
paper~\cite{Han2010}, given a fixed chain length, we focused on the
thermodynamic properties and structure transitions in PAPSs. The
microfluctuation homogenous (MFH) morphology, which corresponds to
the onset of gelation~\cite{Kumar2001,Han2010}, and micelle
morphology were observed. The degrees of aggregations of the above
two morphologies are very different. The volume fraction of stickers
in micelle morphology is much bigger than that in MFH morphology. In
this work, the property of the aggregation in PAPAs is studied using
self-consistent field lattice model. The temperatures at which MFH
and micelle morphologies first appear, which are denoted by
$T_{\mathrm{MFH}}$ and $T_{\rm m}$, respectively, and the average volume
fraction of stickers at the micellar core,
$\langle\phi_{\mathrm{s}}(r_{\mathrm{co}})\rangle$, are calculated. Such calculations
are carried out for different chain lengths and polymer
concentrations. It is found that $T_{\mathrm{MFH}}$, $T_{\mathrm{m}}$ and
$\langle\phi(r_{\mathrm{co}})\rangle$ are dependent on chain length and
polymer concentration, and the relationship of
$\langle\phi_{\mathrm{s}}(r_{\mathrm{co}})\rangle$ with chain length and polymer
concentration is explained in terms of intrachain and interchain
associations.

\section{Theory\label{sec2}}

We consider a system of incompressible PAPSs, where $n_{\mathrm{P}}$
polymers, each of which is composed of $N_{\mathrm{st}}$ segments of sticker monomer type
(attractive group) and $N_{\mathrm{ns}}$ segments of nonsticky monomer type, are
distributed over a lattice. Each sticker monomer is a regularly placed apart
$l$ monomer along the chain backbone. The degree of
polymerization of chain is $N=N_{\mathrm{st}}+N_{\mathrm{ns}}$. In addition to
polymer monomers, $%
n_{\mathrm{h}}$ solvent molecules are placed on the vacant lattice sites.
Sticker, nonsticky monomers and solvent molecules are of the same size
and each occupies one
lattice site. The total number of lattice sites is $N_{\mathrm{L}}$ $=$ $%
n_{\mathrm{h}}+n_{\mathrm{P}}N$. Nearest neighbor pairs of stickers have attractive
interaction $-\epsilon$ with $\epsilon>0$, which is the only
non-bonded interaction in the present system. The interaction energy
is
expressed as:%
\begin{equation}
U=-\frac{\epsilon }{2}\sum_{r}\sum_{{r}^{_{^{\prime }}}}\widehat{%
\phi }_{\mathrm{st}}({r})\widehat{\phi }_{\mathrm{st}}(r'), \label{00}
\end{equation}%
where $\sum_{r}$ means the summation over all the lattice sites ${r}$ and $%
\sum_{{r'}}$ means the summation over the nearest
neighbor sites of ${r}$. $\widehat{\phi
}_{\mathrm{st}}({r})=\sum_{{j}}\sum_{s{\in \mathrm{st}}}\delta
_{{r},{r}_{j,s}}$ is the volume fraction of stickers on site ${r}$, where $j$ and $%
s$ are the indexes of chain and monomer of a polymer, respectively.
$s\in \mathrm{st}$ means that the $s$th monomer belongs to sticker
monomer type. In this simulation, however, instead of directly using the
exact expression of the nearest neighbor interaction for stickers,
we introduce a local
concentration approximation for the non-bonded interaction~\cite{Han2010,Chen2006}. $%
\sum_{{r'}}\widehat{\phi }_{\mathrm{st}}(r')$ in equation~(\ref{00}) is replaced with $z$ $\widehat{\phi }_{\mathrm{st}}({r})$, where
$z$ is the coordination number of the lattice used. Within this
approximation, the
interaction energy is expressed as:%
\begin{equation}
\frac{U}{k_{\mathrm{B}}T}=-\chi \sum_{{r}}\widehat{\phi }_{\mathrm{st}}({r})\widehat{\phi }%
_{\mathrm{st}}({r}),  \label{01}
\end{equation}%
where $\chi $ is the Flory-Huggins interaction parameter in the
solutions, which is equal to $\frac{z}{2k_{\mathrm{B}}T}\epsilon$. We perform the
SCFT calculations in the canonical ensemble, and the field-theoretic
free energy F is defined as follows:
\begin{equation}
\frac{F[\omega _{+},\omega _{-}]}{k_{\mathrm{B}}T}=\sum_{r}\left\{ \frac{1}{4\chi }%
\omega _{-}^{2}(r)-\omega _{+}(r)\right\} -n_{\mathrm{P}}\ln Q_{\mathrm{P}}[\omega
_{\mathrm{st}},\omega _{\mathrm{ns}}]-n_{\mathrm{h}}\ln Q_{\mathrm{h}}[\omega _{\mathrm{h}}],  \label{free0}
\end{equation}%
where $Q_{\mathrm{h}}$ is the partition function of a solvent molecule
subject to the field $\omega _{\mathrm{h}}(r)=$ $\omega _{_{+}}(r)$, which
is defined as $Q_{\mathrm{h}}=\frac{1}{n_{\mathrm{h}}}\sum_{r}\exp\left[ -\ \omega
_{\mathrm{h}}(r)\right]$. $Q_{\mathrm{P}}$ is the partition function of a
noninteraction polymer chain subject to the fields $\omega
_{\mathrm{st}}(r)=\omega _{_{+}}(r)-\omega _{_{-}}(r)$ and $\omega
_{\mathrm{ns}}(r)=\omega _{_{+}}(r)$, which act on sticker and nonsticky
segments, respectively. Following the scheme by Schentiens
and Leermakers~\cite{Leer1988}, $Q_{\mathrm{P}}$ is expressed as $Q_{\mathrm{P}}=\frac{1%
}{N_{\mathrm{L}}}\frac{1}{z}\sum_{r_{\mathrm{N}}}\sum_{{\alpha }_{\mathrm{N}}}G^{\alpha _{\mathrm{N}}}(r,N|1)$%
, where $r_{\mathrm{N}}$ and ${\alpha }_{\mathrm{N}}$ denote the position and
orientation of the $N$th segment of the chain, respectively.
$\sum_{r_{\mathrm{N}}}\sum_{\alpha _{\mathrm{N}}}$ means the summation over all the
possible positions and orientations of the $N$th segment of the
chain. $G^{\alpha _{\mathrm{s}}}(r,s|1)$ is the end segment distribution
function of the $s$th segment of the chain, which is evaluated from
the following recursive relation:
\begin{equation} G^{\alpha
_{\mathrm{s}}}(r,s|1)=G(r,s)\sum_{r_{s-1}^{\prime }}\sum_{\alpha
_{s-1}}\lambda _{r_{\mathrm{s}}-r_{s-1}^{\prime }}^{\alpha _{\mathrm{s}}-\alpha
_{s-1}}G^{\alpha _{s-1}}(r^{\prime },s-1|1), \label{free}
\end{equation}
where $G(r,s)$ is the free segment
weighting factor and is expressed as
\[G(r,s)=
\left\{
  \begin{array}{ll}
    \exp[-\omega
_{\mathrm{st}}(r_{_{\mathrm{s}}})], & \hbox{$s\in \rm st$\,;} \\
    \exp[-\omega _{\mathrm{ns}}(r_{_{\mathrm{s}}})], & \hbox{$s\in
\rm ns$\,.}
  \end{array}
\right.
\]
The initial condition is $G^{\alpha _{1}}(r,1|1)=G(r,1)$ for
all the values of $\alpha _{1}$. In the above expression, the values
of $\lambda _{r_{\mathrm{s}}-r{'}_{s-1}}^{\alpha _{\mathrm{s}}-\alpha _{s-1}}$ depend
on the chain model used. We assume that
\[
\lambda
_{r_{\mathrm{s}}-r{'}_{s-1}}^{\alpha _{\mathrm{s}}-\alpha _{s-1}}=
\left\{
  \begin{array}{ll}
    0, & \hbox{$\alpha _{\mathrm{s}}=\alpha _{s-1}$\,;} \\
    {1}/(z-1), & \hbox{$\mathrm{otherwise}$\,.}
  \end{array}
\right.
\]
This means that the chain is described as a random walk without the
possibility of direct backfolding. Although self-intersections of a
chain are not permitted, the excluded volume effect is sufficiently
taken into account~\cite{Medv2001}. Another end segment distribution
function $G^{\alpha _{\mathrm{s}}}(r,s|N)$ is evaluated from the following
recursive relation: \begin{equation} G^{\alpha
_{\mathrm{s}}}(r,s|N)=G(r,s)\sum_{r_{s+1}^{\prime }}\sum_{\alpha
_{s+1}}\lambda _{r_{s+1}^{\prime }-r_{\mathrm{s}}}^{\alpha _{s+1}-\alpha
_{\mathrm{s}}}G^{\alpha _{s+1}}(r^{\prime },s+1|N),
\end{equation}
with the initial condition $G^{\alpha _{\mathrm{N}}}(r,N|N)=G(r,N)$ for all
the values of $\alpha _{\mathrm{N}}$.

Using the expressions of the end segment distribution functions, the
single-segment probability distribution function $P^{(1)}(r,s)$ and
the two-segment probability distribution function \linebreak $P^{(2)}(r_{1},s_{1};r_{2},s_{2})$ of the  chain can be defined as
follows:

\begin{equation}
P^{(1)}(r,s)=\frac{1}{zN_{\mathrm{L}}Q_{\mathrm{P}}}\sum_{r_{\mathrm{s}}^{\prime }}\sum_{{\alpha }%
_{_{\mathrm{s}}}}\frac{G^{\alpha _{\mathrm{s}}}(r^{\prime },s|1)G^{\alpha
_{\mathrm{s}}}(r^{\prime },s|N)}{G(r^{\prime },s)}\delta _{r_{\mathrm{s}}^{\prime
},{r}}\,,
\end{equation}%
which is the normalized probability that the monomer $s$ of the
chain is on the
lattice site $r$;%
\begin{eqnarray}
P^{(2)}(r_{1},s_{1};r_{2},s_{2}) &=&\frac{1}{zN_{\mathrm{L}}Q_{\mathrm{P}}}%
\sum_{r_{s_{1}}^{^{\prime }}}\sum_{{\alpha }_{_{s_{1}}}}\sum_{r_{s_{2}}^{^{%
\prime }}}\sum_{{\alpha }_{_{s_{2}}}}G^{\alpha _{s_{1}}}(r^{^{\prime
}},s_{1}|1)\delta _{{r}_{s_{1}}^{^{\prime }},{r}_{1}} \nonumber\\
&\times&\mathcal{G}(r^{^{\prime }},s_{1};r^{^{\prime }},s_{2})G^{\alpha
_{s_{2}}}(r^{^{\prime }},s_{2}|N)\delta _{{r}_{s_{2}}^{^{\prime
}},{r}_{2}}
\end{eqnarray}%
and
\[
\mathcal{G}(r,s_{1};r,s_{2})=\sum_{r_{s_{1}+1}}\sum_{\alpha
_{s_{1}+1}}.....\sum_{r_{s_{2}-1}}\sum_{\alpha _{s_{2}-1}}\left\{
\prod_{s=s_{1}+1}^{s_{2}-1}\lambda _{r_{\mathrm{s}}-r_{s-1}}^{\alpha
_{\mathrm{s}}-\alpha _{s-1}}G(r,s)\right\} \lambda
_{r_{s_{2}}-r_{s_{2}-1}}^{\alpha _{s_{2}}-\alpha _{s_{2}-1}}\ \ \ \
\ (\text{for}\ s_{2}>s_{1})
\]%
yield the probability that the monomers $s_{1}$ \emph{and} $s_{2}$ of
the chain are on the lattice sites $r_{1}$ and $r_{2}$,
respectively. It can
be verified that $\sum_{r}P^{(1)}(r,s)=1$, and $%
\sum_{r_{2}}P^{(2)}(r_{1},s_{1};r_{2},s_{2})=P^{(1)}(r_{1},s_{1})$.

Equation~(\ref{free0}) can be considered as the alternative form of the
self-consistent field
free energy functional for incompressible polymer solutions~\cite{Fredr2005}. When a local concentration approximation for the
non-bonded interaction is introduced, the SCFT description of
lattice model for PAPSs presented in this work is basically
equivalent to that of the ``Gaussian thread model'' chain for the
similar polymer solutions~\cite{Fredr2005}. The related details are
presented in~\cite{Han2010}.

 Minimization of the free energy function $F$ with
$\omega _{_{-}}(r)$ and $\omega _{_{+}}(r)$ leads to the following
saddle point equations:
\begin{equation}
\omega _{_{-}}(r)=2\chi \phi _{\mathrm{st}}(r),  \label{scf1}
\end{equation}%
\begin{equation}
\phi _{\mathrm{st}}(r)+\phi _{\mathrm{ns}}(r)+\phi _{\mathrm{h}}(r)=1, \end{equation} where
\begin{equation}
\phi _{\mathrm{st}}(r)=\frac{1}{N_{\mathrm{L}}}\frac{1}{z}\frac{n_{\mathrm{P}}}{Q_{\mathrm{P}}}\sum_{s\in {st}%
}\sum_{\alpha _{\mathrm{s}}}\frac{G^{\alpha _{\mathrm{s}}}(r,s|1)G^{\alpha _{\mathrm{s}}}(r,s|N)}{G(r,s)%
}
\end{equation}
and
\begin{equation}
\phi _{\mathrm{ns}}(r)=\frac{1}{N_{\mathrm{L}}}\frac{1}{z}\frac{n_{\mathrm{P}}}{Q_{\mathrm{P}}}\sum_{s\in {ns}%
}\sum_{\alpha _{\mathrm{s}}}\frac{G^{\alpha _{\mathrm{s}}}(r,s|1)G^{\alpha _{\mathrm{s}}}(r,s|N)}{G(r,s)%
}
\end{equation}
are the average numbers of sticker and nonsticky segments at $r$,
respectively, and \linebreak $\phi_{\mathrm{h}}(r)=\frac{1}{N_{\mathrm{L}}}\frac{n_{\mathrm{h}}}
{Q_{_{\mathrm{h}}}}\exp \left[ -\omega _{\mathrm{h}}(r)\right]$ is the average
numbers of solvent molecules at $r$.

The saddle point is calculated using the pseudo-dynamical evolution
process presented by Fredrickson et al.~\cite{Fred2002}:
\begin{equation} \omega _{_{+}}^{\mathrm{new}}(r)=\omega _{_{+}}^{\mathrm{old}}(r)+\lambda
_{+}(\phi _{\mathrm{st}}(r)+\phi _{\mathrm{ns}}(r)+\phi _{\mathrm{h}}(r)-1),
\end{equation}
\begin{equation} \omega _{_{-}}^{\mathrm{new}}(r)=\omega _{_{-}}^{\mathrm{old}}(r)+\lambda
_{-}(\phi _{\mathrm{st}}(r)-\frac{\omega _{_{-}}(r)}{2\chi }).
\end{equation}

The calculation is initiated from appropriately randomly-chosen fields $%
\omega_{_{+}}(r)$ and $\omega _{-}(r)$, and stopped when the change
of free energy $F$ between two successive iterations is reduced to
the needed precision. The resulting configuration is taken as a
saddle point configuration. By comparing the free energies of the saddle point
configurations obtained from different initial fields, the relative
stability of the observed morphologies can be assessed.

\section{Result and discussion\label{sec3}}

In our studies, the properties of associating polymers depend on
three tunable molecular parameters: $\chi$ (The Flory-Huggins
interaction parameter), $N$ (Chain length) and $l$ (The number of
nonsticky monomers between two neighboring stickers along the
backbone, $l$ equals 9 in this paper). The calculations are
performed in three-dimensional simple cubic lattice with periodic
boundary condition. The aggregation behavior is first sketched
using the lattice with the size $N_{\mathrm{L}}=26^{3}$.
Then, the obtained results are verified using larger size lattices. The results
presented below are obtained from the lattice with
$N_{\mathrm{L}}=40^{3}$. Three different morphologies, i.e., the
homogenous, micro-fluctuation homogenous and micelle morphologies,
are observed in PAPSs. By comparing the relative stability of the
observed states, the phase diagram is constructed.

%
\begin{figure}[!ht]
\centerline{\includegraphics[width=0.55\textwidth]{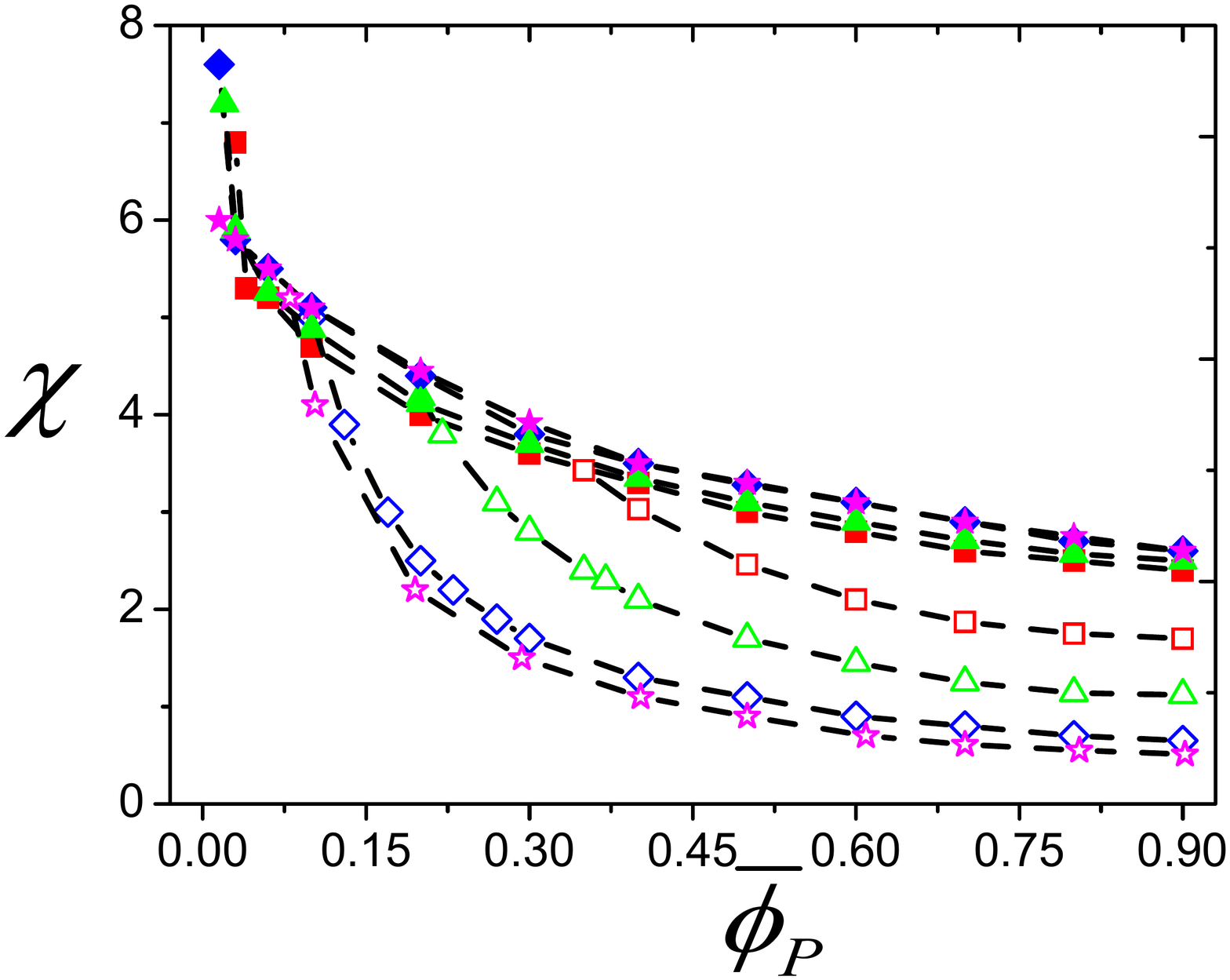}}
\caption{(Color online) The phase diagram for the systems with different chain
lengths $N$. The boundaries of MFH and micelle morphologies are
obtained. The red open and solid squares, green open and solid
triangles, blue open and solid diamonds, magenta open and solid
pentagons correspond to the boundaries of MFH and micelle
morphologies for $N=21, 41, 81,101$, respectively.\label{Nchange}}
\end{figure}
%

%
\begin{figure}[!ht]
\centerline{\includegraphics[width=0.55\textwidth]{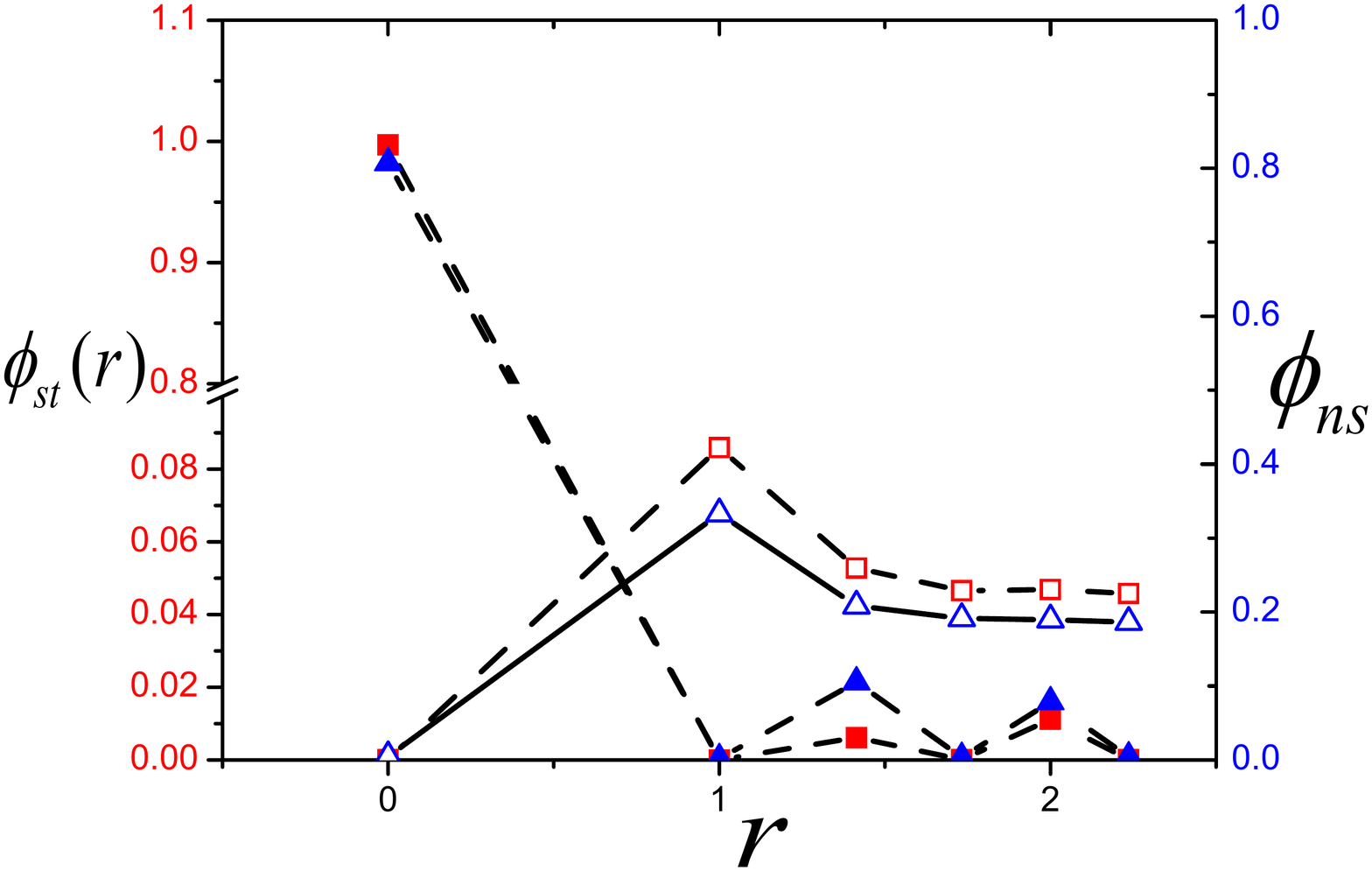}}
\caption{(Color online) The variations of the volume fractions of sticker and
nonsticker components of polymers in micelle morphology  with $r$
when $\bar{\phi}_{\mathrm{P}}=0.1$ and $\chi=5.5$ in the systems with $N=21$
and $101$. r is the distance from the sticker-rich core. The red
solid and open squares, blue solid and open triangles denote the
volume fractions of stickers and nonstickers for $N=101$ and $21$,
respectively.\label{micecom}}
\end{figure}
%

Figure~\ref{Nchange} shows the phase diagram of the systems with
different chain lengths $N$. In this study, when $N$ is changed, only
MFH morphology and micelles are observed as inhomogeneous
morphologies. The structural morphology of MFH morphology does not
change, and micellar shape remains spherelike (see figure~\ref{micecom}). For $N=21$, the $\chi$ value on micellar boundary
($\sim{1}/{T_{\mathrm{m}}}$) increases with decreasing $\bar{\phi}_{\mathrm{P}}$.
Approaching to critical micelle concentration
($\bar{\phi}_{\mathrm{CMC}}=0.04$)\footnote{Critical micelle concentration (CMC) is generally considered as the minimum
concentration of the micellar appearance. In this study, the
concentration which corresponds to the $\chi$ abrupt increase on
micellar boundary is regarded as CMC, which is denoted by
$\bar{\phi}_{\mathrm{CMC}}$. When $N=21$, $\bar{\phi}_{\mathrm{CMC}}=0.04$. When
$N$ is increased $\bar{\phi}_{\mathrm{CMC}}$ decreases. For $N=101$, the
$\bar{\phi}_{\mathrm{CMC}}$ is not observed till $\bar{\phi}_{\mathrm{P}}=0.015$.}, micellar boundary abruptly
becomes steep. The $\chi$ value on MFH boundary
($\sim{1}/{T_{\mathrm{MFH}}}$) also goes up with the decrease in
$\bar{\phi}_{\mathrm{P}}$, which resembles the behavior on the micellar
boundary. The critical MFH concentration
($\bar{\phi}_{\mathrm{CFC}}=0.37$) is much higher than that of micellar
morphology. Therefore, the MFH boundary intersects with the micellar
boundary at $\bar{\phi}_{\mathrm{CFC}}$. When $N$ is increased, at fixed
$\bar{\phi}_{\mathrm{P}}$, the $\chi$ value on micellar boundary shifts
slightly to larger value, and the $\chi$ value on MFH boundary
decreases markedly, which is different from that on micellar
boundary. $\bar{\phi}_{\mathrm{CFC}}$  also decreases with the increase in
$N$.


\begin{figure}[!h]
\centering
\includegraphics[width=0.55\textwidth]{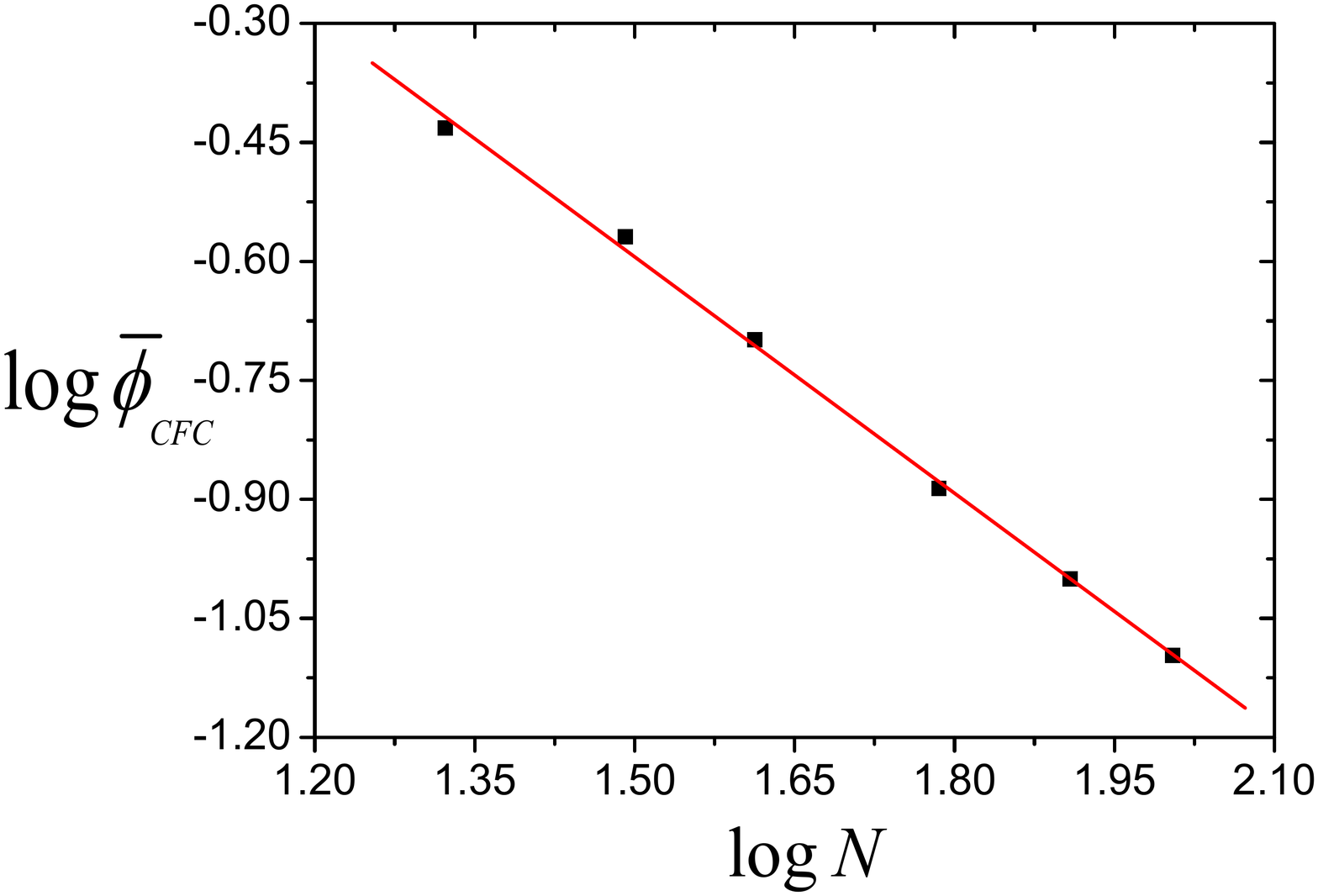}\hspace{1cm}
\caption{The logarithm of critical MFH concentration
$\bar{\phi}_{\mathrm{CFC}}$ a function of the logarithm of chain length
$N$, which fulfils a linear-fitting relationship with the slope
equaling -1. \label{triphase}}
\end{figure}

\begin{figure}[!h]
\centering
\includegraphics[width=0.55\textwidth]{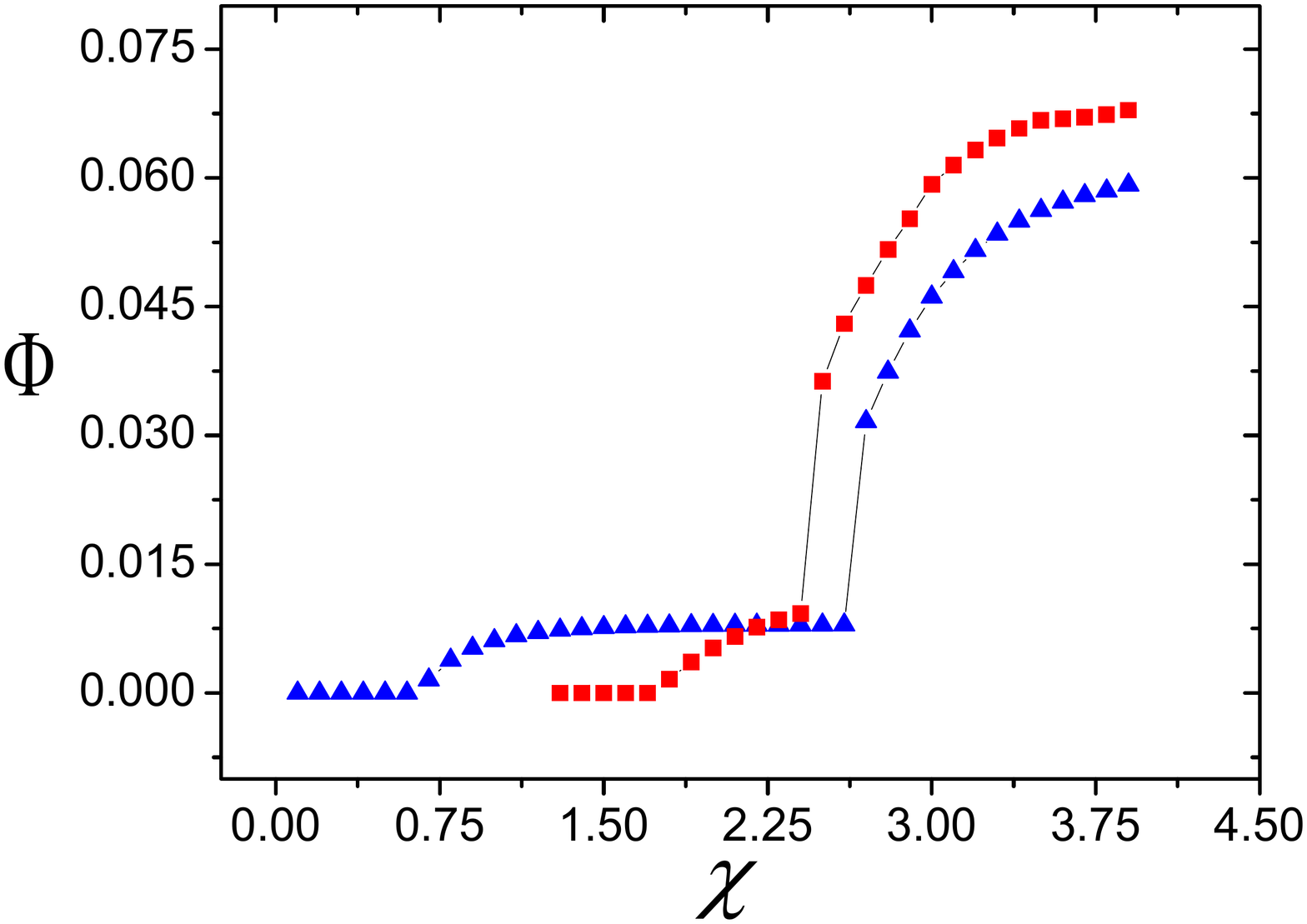}
\caption{(Color online) The variation of the order-parameter-type variable $\Phi$
with $\chi$ near the HS-MFH  and MFH-micelle transition points at
$\bar{\phi}_{\mathrm{P}}=0.8$ in the systems with different chain length
$N$. The red squares and blue triangles correspond to $N=21$ and
$N=81$, respectively. With the increase in $\chi$, there are the
appearances of two nonzero steps on the above two curves, which
denote the HS-MFH and MFH-micelle transitions, respectively.
\label{phi}}
\end{figure}

Figure~\ref{triphase} shows the variations of the logarithm of
$\bar{\phi}_{\mathrm{CFC}}$ with the logarithm of $N$. It is seen that the
straight line with a slope equaling $-1$ fits the results for all
chain lengths considered in this study. The appearance of MFH
morphology is considered as the onset of
gelation~\cite{Kumar2001,Han2010}. Therefore, critical MFH
concentration $\bar{\phi}_{\mathrm{CFC}}$ should correspond to the
critical gelation concentration $C_*$, which is analyzed in terms of
the concept of chain overlap. The critical gelation concentration
$C_*$, at which quasi-ideal coils begin to overlap, the pervaded
volume of one another is related to the chain length as
$N=KC_*^{-2}$, where K is a
constant~\cite{Daou1976,Daou1975,Klei1978}. The corresponding slope
of the critical concentration $C_*$ at quasi-ideal coil is $-2$.
Similarly, at the excluded volume chain, the corresponding quantity
is $-5/4$. The slope of fitting straight line at associative polymer
chain is the smallest of the above three cases. It is demonstrated
that the physically associating polymer chain in solution should be
elongated compared with the excluded volume chain and quasi-ideal
coil.

In order to measure the degree of aggregation of stickers, an
order-parameter-type variable~\cite{Han2010} is used in this work. It
is expressed as:
\begin{equation}
\Phi (\left\{ \phi _{\mathrm{st}}(r)\right\} )=\frac{1}{N_{\mathrm{L}}}\sum_{r}(\phi _{\mathrm{st}}(r)-%
\bar{\phi}_{\mathrm{st}})^{2}=\frac{1}{N_{\mathrm{L}}}\sum_{r}\phi _{\mathrm{st}}^{2}(r)-\bar{\phi}%
_{\mathrm{st}}^{2}\,,
\end{equation}%
where $\Phi $ is determined by the distribution of the volume
fraction of stickers $\left\{ \phi _{\mathrm{st}}(r)\right\} $, which is a
function of $\chi $ and $\bar{\phi}_{\mathrm{P}}$. For homogenous
solutions (HS), $\Phi $ is equal to zero. Figure~\ref{phi} shows the
variations of $\Phi $ in MFH and micelle morphologies  with an
increasing $\chi$ when $\bar{\phi}_{\mathrm{P}}=0.8$ near HS-MFH and
MFH-micelle transition points, in the systems with $N=21$ and
$N=81$. It is seen that the degree of the aggregation of stickers
increases with an increasing $\chi $ at fixed $N$. Compared with
$N=81$, although the value of $\Phi $ at fixed $\chi$ above
MFH-micelle transition point at $N=21$ is larger, the variations of
$\Phi $ on the above two cases with $\chi$ are similar. However,
near HS-MFH transition point, the variation of $\Phi $ with $\chi$
at $N=21$ is obviously different from that of $N=81$. The value of
$\Phi$ at $N=21$ increases more rapidly than that of $N=81$ when
$\chi$ is increased. The heat capacity $C_{\mathrm{V}}$ is proportional to
the first derivative of $\Phi $ with respect to
temperature~\cite{Han2010,Dudo1999,Ken83}. When $N$ is increased, the
maximum of the peak of $C_{\mathrm{V}}$ decreases, and the corresponding
half-width increases near the HS-MFH transition. The shape of the
peak of $C_{\mathrm{V}}$ near MFH-micelle transition point does not
practically change with the increase in $N$ (not shown). It is demonstrated that
the change of $N$ has a greater effect on HS-MFH transition than
that on MFH-micelle transition.

\begin{figure}[h]
\centering
\includegraphics[width=0.55\textwidth]{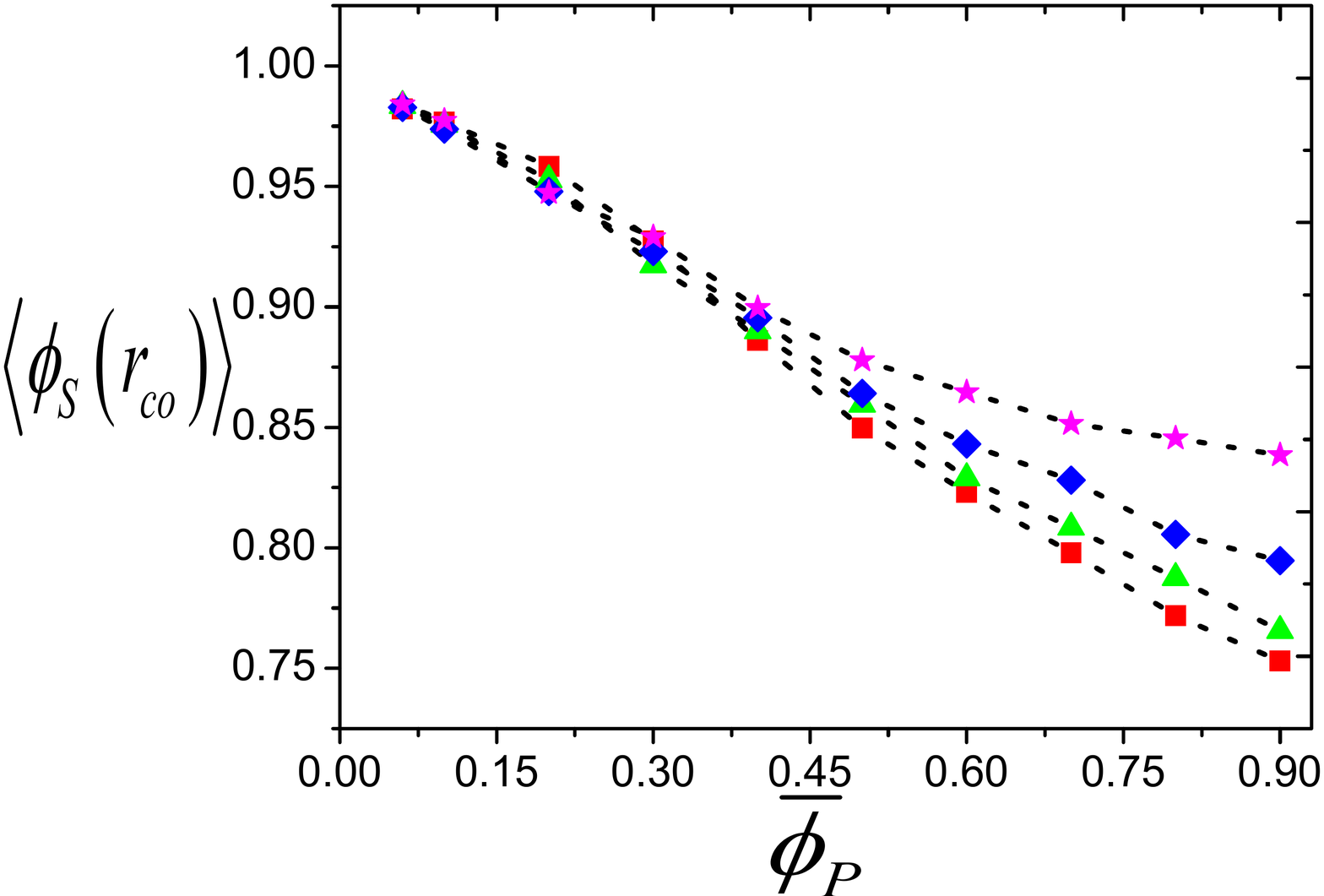}
\caption{(Color online) The average volume fractions of stickers at micellar core
as a function of $\bar{\phi}_{\mathrm{P}}$ in micellar boundary systems with
different chain lengths. The red squares, green triangles, blue
diamonds and magenta pentagons correspond to the boundaries of
micelle morphology for $N=21, 41, 81,101$, respectively.
\label{msphi}}
\end{figure}

The aggregation number of micelles, which is used to account for
some properties, for example, the rheological behavior of
associative polymers, is a major focus in  many experimental~\cite{Elli2003,Voro2001, Sere1998} and
theoretic~\cite{Meng2005} studies. In this paper, the average volume
fraction of stickers at micellar cores
$\langle\phi_{\mathrm{s}}(r_{\mathrm{co}})\rangle$ is similar to the average
aggregation number of the micelle. Figure~\ref{msphi} shows the
variations of $\langle\phi_{\mathrm{s}}(r_{\mathrm{co}})\rangle$ on micellar boundary
with $\bar{\phi}_{\mathrm{P}}$ and $N$ in the systems. Being given a fixed chain
length, when $\bar{\phi}_{\mathrm{P}}$ is increased,
$\langle\phi_{\mathrm{s}}(r_{\mathrm{co}})\rangle$ decreases. When $N$ is increased,
at fixed $\bar{\phi}_{\mathrm{P}}$, the variation of
$\langle\phi_{\mathrm{s}}(r_{\mathrm{co}})\rangle$ with $\bar{\phi}_{\mathrm{P}}$, does
not practically change for $\bar{\phi}_{\mathrm{P}}\leqslant 0.4$, on the other hand, when
$0.4<\bar{\phi}_{\mathrm{P}}\leqslant 0.9$, $\langle\phi_{\mathrm{s}}(r_{\mathrm{co}})\rangle$
increases. The decreasing tendency of
$\langle\phi_{\mathrm{s}}(r_{\mathrm{co}})\rangle$  with $\bar{\phi}_{\mathrm{P}}$ becomes
gentle with the increase in $N$ at high polymer concentrations. It
is shown that the effect of the increase in $N$ on
$\langle\phi_{\mathrm{s}}(r_{\mathrm{co}})\rangle$ is related to
$\bar{\phi}_{\mathrm{P}}$.

In order to interpret the effects of $N$ and $\bar\phi_{\mathrm{P}}$ on the
aggregation of stickers in micelle morphology, we evaluate the
probabilities that a polymer chain forms intrachain and interchain
associations in the system using an approach similar to the one
presented in Refs.~\cite{Bras2009,Mats1999,Han2010}. We suppose that
there are no other sticker aggregates in the micellar system except
the micellar cores because the volume fraction of stickers at
micellar core approaches $1$, and that of the rest of the system
is less than $10^{-1}$. A sticker in a particular chain can form
an intrachain association, as well as interchain association. Ignoring
the probabilities that more than two stickers of a definite chain
are attached to a micellar core, the conditional probability that
the sticker $s_{1}$ concerns the intrachain association,
provided that the sticker $s_{1}$ is at the micelle core $r_{\mathrm{co}}$,
can be expressed as:

\begin{equation}
p_{\mathrm{loop}}(r_{\mathrm{co}},s_{1})=\frac{1}{P^{(1)}(r_{\mathrm{co}},s_{1})}\sum_{{s_{2}\in st,}%
s_{2}\neq s_{1}}P^{(2)}(r_{\mathrm{co}},s_{1};r_{\mathrm{co}},s_{2}),
\end{equation}%
where $\sum_{{s_{2}\in st,s_{2}\neq s_{1}}}$ means the summation
over all
the stickers of a polymer chain except the $s_{1}$th one, $%
P^{(1)}(r_{\mathrm{co}},s_{1})$ and $P^{(2)}(r_{\mathrm{co}},s_{1};r_{\mathrm{co}},s_{2})$ are
the single-segment and two-segment probability distribution
functions of a chain, respectively. Then, $1-p_{\mathrm{loop}}(r_{\mathrm{co}},s_{1})
$ is the conditional probability that the sticker $s_{1}$ is linked
with those
belonging to other chains when the sticker $s_{1}$ is at $r_{\mathrm{co}}$, and%
\[
\mathbf{P}_{lk}(s_{1})=\sum_{r_{\mathrm{co}}}P^{(1)}(r_{\mathrm{co}},s_{1})\cdot
\left\lceil 1-p_{\mathrm{loop}}(r_{\mathrm{co}},s_{1})\right\rceil
\]%
is the probability that a sticker $s_{1}$ of a chain is related to
interchain association, where $\sum_{r_{\mathrm{co}}}$ means the summation
over all the micellar cores of the system. The summation of
$\mathbf{P}_{lk}(s_{1})$ over all the stickers in a chain, $\langle
n_{_{lk}}\rangle =\sum_{s_{1},s_{1}\in st}\mathbf{P}_{lk}(s_{1})$,
can be viewed as the average sticker number from a particular
polymer chain linked with other chains by sticker aggregates. The
total number of stickers participating in interchain association in
the system is expressed as $n_{\mathrm{inte}}=n_{\mathrm{P}}\langle
n_{_{lk}}\rangle$. The total number of stickers belonging to
intrachain association is defined as $n_{\mathrm{intr}}
=(1/N_{\mathrm{st}})[\sum_{s_{1},s_{1}\in
st}\{(1/N_{\mathrm{m}})\{\sum_{{r_{\mathrm{co}}}}p_{\mathrm{loop}}(r_{\mathrm{co}},s_{1})\}\}]\cdot\langle\phi_{\mathrm{s}}(r_{\mathrm{co}})\rangle
N_{\mathrm{m}}$, where $N_{\mathrm{m}}$ denotes the micellar number in the system.

\begin{figure}[h]
\centering
\includegraphics[width=0.55\textwidth]{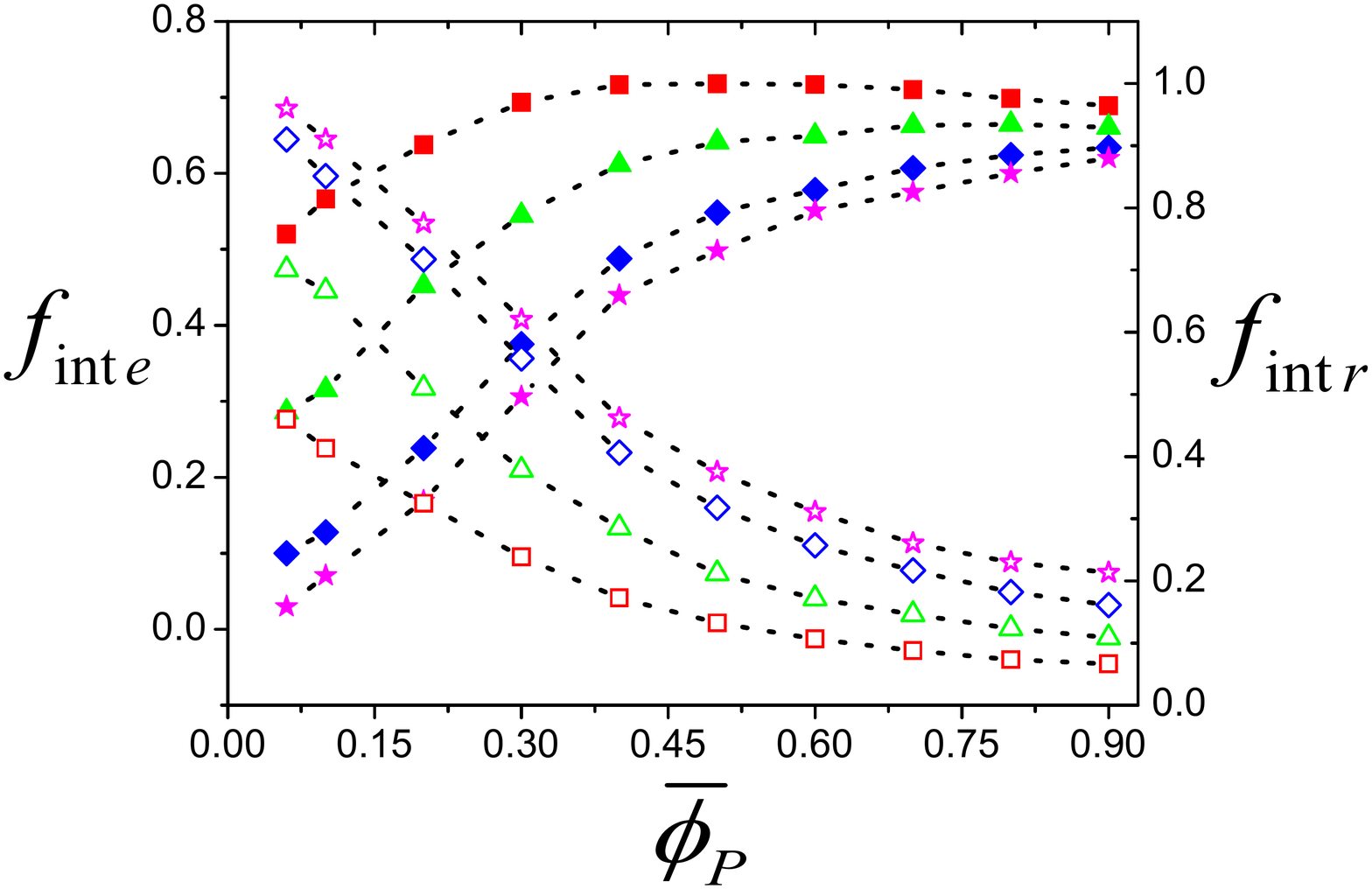}
\caption{(Color online) The variations of the average fractions of intrachain and
interchain associations at a micellar core with $\bar{\phi}_{\mathrm{P}}$ in
the same systems presented in figure~\ref{msphi}, denoted by
$f_{\mathrm{intr}}$ and $f_{\mathrm{inte}}$, respectively. The red open and solid
squares, green open and solid triangles, blue open and solid
diamonds, magenta open and solid pentagons correspond to $f_{\mathrm{intr}}$
and $f_{\mathrm{inte}}$ for $N=21, 41, 81, 101$, respectively.\label{loop}}
\end{figure}

Figure~\ref{loop} shows the variations of the average fractions of
intrachain and interchain associations within a micelle, denoted by
$f_{\mathrm{intr}}=n_{\mathrm{intr}}/N_{\mathrm{m}}$ and $f_{\mathrm{inte}}=n_{\mathrm{inte}}/N_{\mathrm{m}}$, respectively,
with $\bar{\phi}_{\mathrm{P}}$ and $N$ in the systems identical with those
presented in figure~\ref{msphi}. Being given a fixed chain length, when
$\bar{\phi}_{\mathrm{P}}$ is increased, $f_{\mathrm{intr}}$ does decrease
markedly first, and then the decreasing tendency of $f_{\mathrm{intr}}$ becomes
gentle when $\bar{\phi}_{\mathrm{P}}$ exceeds some value. Except for the
case of $N=21$, $f_{\mathrm{inte}}$  at fixed $N$  increases markedly first,
and then increases gently with the increase in $\bar{\phi}_{\mathrm{P}}$,
which is contrary to $f_{\mathrm{intr}}$. At $N=21$, $f_{\mathrm{inte}}$ increases to
$\bar{\phi}_{\mathrm{P}}=0.6$, then decreases slightly with the increase
in $\bar{\phi}_{\mathrm{P}}$. When $N$ is increased, $f_{\mathrm{intr}}$ at fixed
$\bar{\phi}_{\mathrm{P}}$ increases, and corresponding $f_{\mathrm{inte}}$
decreases. When $\bar{\phi}_{\mathrm{P}}$ is increased the differences of
$f_{\mathrm{intr}}$ and $f_{\mathrm{inte}}$ in various $N$ become small. In other
words, the effect of chain length on $f_{\mathrm{intr}}$ and $f_{\mathrm{inte}}$ at low
concentrations  is more pronounced than that at high concentrations.

The relative magnitude of $f_{\mathrm{intr}}$ and $f_{\mathrm{inte}}$ at fixed $N$ is
dependent on $\bar{\phi}_{\mathrm{P}}$. Except for $N=21$, when
$\bar{\phi}_{\mathrm{P}}$ is low, $f_{\mathrm{inte}}$ at fixed $N$ is smaller than
corresponding $f_{\mathrm{intr}}$. For $N=101$, $f_{\mathrm{intr}}$ equals 1, and
corresponding $f_{\mathrm{inte}}$ equals =0 at $\bar{\phi}_{\mathrm{P}}=0.015$ and
$\chi=6.0$. The micelle is absolutely aggregated by intrachain
association. When $\bar{\phi}_{\mathrm{P}}$ exceeds some value, which is
related to $N$, $f_{\mathrm{inte}}$ at fixed $N$ is larger than
corresponding $f_{\mathrm{intr}}$. When $N=21$, $f_{\mathrm{inte}}$ is always larger
than $f_{\mathrm{intr}}$ in the considered range of $\bar{\phi}_{\mathrm{P}}$.

\looseness=1In light of the variations of
 $\langle\phi_{\mathrm{s}}(r_{\mathrm{co}})\rangle$ with $\bar{\phi}_{\mathrm{P}}$ and $N$ in the systems,
 the effect of chain length on aggregation of stickers is more pronounced at high concentrations.
However, seen from respective value of $f_{\mathrm{intr}}$ and $f_{\mathrm{inte}}$ at
fixed $\bar{\phi}_{\mathrm{P}}$, the increase of chain length is more
favorable to the change of intrachain and interchain associations at
low concentrations, which is different from the above conclusions
drawn in light of $\langle\phi_{\mathrm{s}}(r_{\mathrm{co}})\rangle$. It is shown that
[see figure~\ref{loop}], being given a fixed chain length, although
$f_{\mathrm{intr}}$ is smaller than the corresponding $f_{\mathrm{inte}}$ at high
concentrations, the decrease of $\langle\phi_{\mathrm{s}}(r_{\mathrm{co}})\rangle$
with the increase in $\bar{\phi}_{\mathrm{P}}$ is similar to that of
$f_{\mathrm{intr}}$, instead of $f_{\mathrm{inte}}$, in the considered range of
$\bar{\phi}_{\mathrm{P}}$. When $N$ is increased, $f_{\mathrm{intr}}$ at fixed
$\bar{\phi}_{\mathrm{P}}$ goes up. It is notable that the effect of chain
length on $f_{\mathrm{intr}}$  at low concentrations is much more pronounced than
that at high concentrations. However, the increase of $f_{\mathrm{intr}}$ at
fixed $\bar{\phi}_{\mathrm{P}}$ does not result in the dramatical change
of $\langle\phi_{\mathrm{s}}(r_{\mathrm{co}})\rangle$ at low concentrations. At high
concentrations ($\bar{\phi}_{\mathrm{P}}>0.4$), on the contrary,
$\langle\phi_{\mathrm{s}}(r_{\mathrm{co}})\rangle$ with a big chain length has a
larger value. It is reasonable that $\langle\phi_{\mathrm{s}}(r_{\mathrm{co}})\rangle$
is also affected by $f_{\mathrm{inte}}$. When $\bar{\phi}_{\mathrm{P}}$ is low,
$f_{\mathrm{inte}}$ at fixed $\bar{\phi}_{\mathrm{P}}$ with a long chain length is
much smaller than that with a short chain. Therefore, the increase
of $f_{\mathrm{intr}}$ at fixed $\bar{\phi}_{\mathrm{P}}$ is canceled by the
decrease of corresponding $f_{\mathrm{inte}}$, thus increasing in $N$. With the
increase in $\bar{\phi}_{\mathrm{P}}$, $f_{\mathrm{inte}}$ at fixed
$\bar{\phi}_{\mathrm{P}}$ goes up markedly. The difference of $f_{\mathrm{inte}}$
among different $N$ becomes small. The contribution of $f_{\mathrm{intr}}$ to
$\langle\phi_{\mathrm{s}}(r_{\mathrm{co}})\rangle$ resulting from the increase in $N$
begins to appear. It is demonstrated that the effects of chain
length and polymer concentration on $\langle\phi_{\mathrm{s}}(r_{\mathrm{co}})\rangle$
do interact each other. The increase of chain length is favorable to
intrachain association, and retains interchain association, which
is contrary to the effect of the increase in $\bar{\phi}_{\mathrm{P}}$.
When $\bar{\phi}_{\mathrm{P}}$ is increased, the effect of chain length on
intrachain and interchain associations is weakened. In other words,
the polymer concentration and chain length simultaneously control
the formations of intrachain and interchain associations.

\section{Conclusion and summary\label{sec4}}

Using the self-consistent field lattice model, the effects of
polymer concentration $\bar{\phi}_{\mathrm{P}}$ and chain length $N$ on
aggregation in physically associating polymer solutions are studied.
When $N$ is changed, only two inhomogenous aggregates, i.e., the MFH
and micelle morphologies, are observed in PAPSs. When
$\bar{\phi}_{\mathrm{P}}$ is decreased, being given a fixed $N$, the $\chi$
values on MFH and micellar boundaries increase. At fixed
$\bar{\phi}_{\mathrm{P}}$, the increase in $N$ remarkably decreases the
$\chi$ value on MFH boundary, but slightly increases the $\chi$
value on micellar boundary. The logarithm
 of critical MFH concentration as a function of the logarithm of $N$ fulfils a fitting
straight line with a slope equaling -1, which demonstrates that the
associating polymer chain in solution should be elongated compared
with an excluded volume chain. Furthermore, on micellar boundary,
the average volume fraction of stickers at a micellar core,
$\langle\phi_{\mathrm{s}}(r_{\mathrm{co}})\rangle$, which is similar to the average
aggregation number, decreases at fixed $N$ when $\bar{\phi}_{\mathrm{P}}$
is increased. With the increase in $N$, on the other hand,
$\langle\phi_{\mathrm{s}}(r_{\mathrm{co}})\rangle$, at fixed $\bar{\phi}_{\mathrm{P}}$,
does not practically change when $\bar{\phi}_{\mathrm{P}}\leqslant 0.4$. When
$\bar{\phi}_{\mathrm{P}}>0.4$, on the other hand, the increase in $N$
causes the increase of $\langle\phi_{\mathrm{s}}(r_{\mathrm{co}})\rangle$ at fixed
$\bar{\phi}_{\mathrm{P}}$. There is found a decreasing tendency of
variation of $\langle\phi_{\mathrm{s}}(r_{\mathrm{co}})\rangle$ at fixed $N$, with
$\bar{\phi}_{\mathrm{P}}$ is determined by intrachain association, instead
of interchain association, and the magnitude of
$\langle\phi_{\mathrm{s}}(r_{\mathrm{co}})\rangle$ with different $N$ is affected by
intrachain and interchain associations. At fixed
$\bar{\phi}_{\mathrm{P}}$, when the difference of the contribution of
interchain association to $\langle\phi_{\mathrm{s}}(r_{\mathrm{co}})\rangle$ between
different $N$ goes down to some extent, the corresponding effect of
intrachain association on $\langle\phi_{\mathrm{s}}(r_{\mathrm{co}})\rangle$ begins to
appear.

The architectural parameters of polymer are important to the
properties of PAPSs. In this paper, only the effect of chain length
of the polymer is investigated. When the other parameters, for
example, the nonsticky monomer number between two neighboring
stickers $l$, are changed, the aggregation behavior is different.
When $l$ is increased,  the $\chi$ values on MFH and micellar
boundaries rise. The increase of $l$ has a different effect on the
specific heat peaks for HS-MFH and MFH-micelle transitions. Systemical
studies are presented in our subsequent work. Our
calculations are within the mean field framework, which may also be
taken as a starting point to the further study, i.e., the direct
calculation of the partition function of the system using complex
Langevin simulations~\cite{Fredr2005}.

\section*{Acknowledgements}
This research is supported by the Innovation Fund of Inner Mongolia University of Science and Technology (Grant No. 2010NC065) the High Performance Computers of
Inner Mongolia University of Science and Technology.



\newpage

\ukrainianpart

\title{Вплив концентрації полімера  і довжини ланцюга  \\ на агрегацію у фізично асоційованих \\розчинах полімерів}

\author{К.Ґ. Ган\refaddr{label1}, К.-Ф. Жанг\refaddr{label1}, Й.-Г. Ма\refaddr{label1}, С.-К. Жан\refaddr{label2}, Й.-Б. Ґуан\refaddr{label2}}
\addresses{
\addr{label1} Школа математики, фізики і біотехнологій, Університет  науки і технологій Внутрішньої Монголії, 014010 Баоту, Китай
\addr{label2} Фізичний факультет, Університет Джілін, 130021 Чанґчун, Китай
}
%
%
%

\makeukrtitle

\begin{abstract}
\tolerance=3000%
Використовуючи ґраткову модель самоузгодженого поля, вивчається вплив концентрації полімера  і довжини
ланцюга   на агрегацію в асоціативних полімерних розчинах. У системах з різною довжиною ланцюга спостерігається тільки дві неоднорідні морфології, а саме мікрофлуктуаційна однорідна (MFH) і міцелярна морфології.
Температури, при яких вище згадані дві морфології виникають вперше і позначаються як $T_{\mathrm{MFH}}$ і $T_{\mathrm{m}}$, відповідно,
є незалежними від концентрації полімера і довжини ланцюга.
Зміна логарифма критичної концентрації MFH зі зміною  логарифма довжини ланцюга задовільняє співвідношенню
лінійного допасування з нахилом рівним  $-1$. Крім того, зміна середньої об'ємної фракції стікерів при міцелярному корі (AVFSM)
з концентрацією полімера і довжиною ланцюга сфокусована в системі при  $T_{\mathrm{m}}$. Знайдено шляхом розрахунків, що  вище згадана
поведінка AVFSM пояснюється в термінах інтраланцюгових та інтерланцюгових асоціацій.
\keywords концентрація, довжина ланцюга, агрегація, асоціативний полімер
\end{abstract}

\end{document}